\documentclass[preprint]{aastex}

\shorttitle{Magnetar  4U 0142+61}
\shortauthors{A.~A. ~Ershov and Yu.~P. ~Shitov}

\begin{document}

\title{ A search for pulsed radio emission from anomalous X-ray pulsar 4U~0142+61
at the frequency of 111~MHz}

\author{Alexander A. Ershov \altaffilmark{1,2} and Yurii P. Shitov \altaffilmark{1,3}}

\altaffiltext{1}{Pushchino Radio Astronomy Observatory, Astro Space
Center, Lebedev Physical Institute, Russia}

\altaffiltext{2}{E-mail:  ershov@prao.ru}

\altaffiltext{3}{Deceased 2007 January 20}

\begin{abstract}
We have searched for pulsed radio emission from magnetar 4U
0142+61 at the frequency of 111 MHz. No pulsed signal was detected
from this source. Upper limits for mean flux density are
0.9~-~9~mJy depending on assumed duty cycle (.05~-~.5) of the
pulsar.
\end{abstract}

%\bigskip

\section{Introduction}

4U~0142+61 is an anomalous X-ray pulsar (AXP) with 8.7 seconds
period. At period derivative $\dot P = 2 \times 10^{-12}$ magnetic
field on a surface of a neutron star equals to $1.3 \times
10^{+14}$~G - typical value for magnetars. The pulsar can be seen
in hard X-ray (\cite{den06}, \cite{kui06}), optical (\cite{ker02}),
and infrared (\cite{hul04}, \cite {wan06}) bands. We searched for pulsed
radio emission of this AXP at the frequency of 111~MHz.

\section{Observations and Data Reduction}

The observations were performed from December 2000 through March
2007 with the Large Phase Array (BSA) radio telescope at Pushchino
Radio Astronomy Observatory with an effective area at zenith of
about 15,000 square meters. One linear polarization was received.
We used 128-channel receiver with a channel bandwidth of 20~kHz
and a center frequency of 110.59~MHz. The observations were
carried out in the mode of recording individual pulses. The
sampling interval was 25.4~ms at the receiver time constant $\tau
= 30$~ms. Since BSA radio telescope is a transit one, the duration
of one observing session is limited to 6.7~min. A total of 540
observational sessions containing 25,920 periods of the pulsar was
carried out. Since the middle of 2004 antenna was calibrated by
observations of the 3C~452 source, flux density of which at 111~MHz
is considered as 91~Jy.

At primary processing of day observation session, a mean value was
deduced from time-series in each frequency channel and a result
was divided by mean square deviation of the channel. Then records
were reviewed to reveal interference, namely: records of all
channels were averaged without compensation for dispersion delay
(as ground interferences have no dispersion delay) and if
interference was revealed (with signal to noise ratio of seven or
higher) then corresponding values were substituted with zero at
all channels. Further: folding, that is summation of periods in
record of each channel, was performed; at that, period meaning for
a special day of observations was calculated on the base of recent
ephemeris of Dib~et~al.~(2006). And finally: compensation of
dispersion delay was performed for each channel; at that,
dispersion measure was searched in the range from 0 to 200~$\rm
pc~cm^{-3}$ with spacing of $2~\rm pc~cm^{-3}$. The proposed
distance (3.6~kpc) to 4U~0142+61 corresponds to dispersion measure
$DM~=~96~\rm pc~cm^{-3}$. No statistically meaning radio emission
was found in any series of observations.

\section{Results}

To improve the sensitivity of the search, all 540 observational
sessions were averaged together by time reference in accordance
with ephemeris of Dib~et~al.~(2006) and with above mentioned
searches of dispersion measure. We did not reveal any significant
radio emission at this processing either. Examples of resulting
means (for all 540 sessions) of pulse profiles for a number of
dispersion measure values are presented at the figure. Value of
upper limit ($5 \sigma$) for a peak flux density equals to 18~mJy.
A correspondent value of the mean (by period) upper limit is
within 0.9 to 9~mJy range depending on a assumed (.05 to .5
of period) pulse duration.

\section{Conclusion}

The search of pulsed radio emission from anomalous X-ray pulsar
(magnetar) 4U~0142+61 at the frequency of 111~MHz give no positive
results.  Upper limits for mean flux density are 0.9~-~9~mJy
depending on assumed duty cycle (.05 - .5) of the pulsar.

\section{Acknowledgements}

We are grateful to the staff of the Pushchino Radio Astronomy
Observatory for help in preparation and performing of
observations. This paper was submitted as a poster for the 40
Years of Pulsars Conference. But because of non-arrival of the
author the conference organizers would not let another person to
mount this poster, unfortunately.

%%\clearpage

\begin{figure}
\center
\includegraphics[scale=1.00]{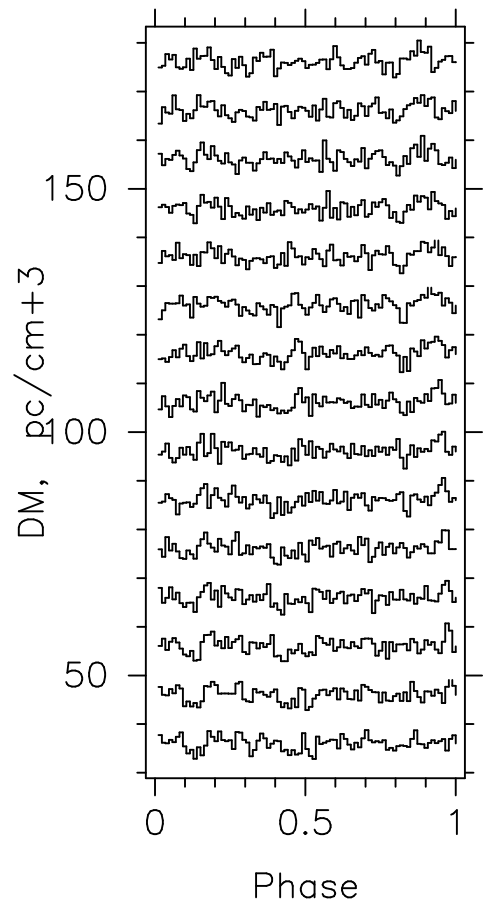}
%\epsscale{.70}
%\plotone{0142_fig.ps}
\caption{
The average profiles of 4U~0142+61 at the frequency of
111~MHz for a number of dispersion measure values.
\label{fig1}}
\end{figure}

\end{document}